\documentclass[12pt]{article}
\usepackage{epsfig}

\def\beq{\begin{equation}}
\def\eeq#1{\label{#1}\end{equation}}
\def\eeqn{\end{equation}}

\def\beqa{\begin{eqnarray}}
\def\eeqa#1{\label{#1}\end{eqnarray}}
\def\eeqan{\end{eqnarray}}

\let\bar=\overbar

\def\Dslash{\not{\hbox{\kern-4pt $D$}}}
\def\dslash{\not{\hbox{\kern-2pt $\del$}}}

\def\msb{{\bar{\ssstyle M \kern -1pt S}}}

\def\Title#1{\begin{center} {\Large {\bf #1} } \end{center}}

\begin{document}

\Title{\boldmath $P_5'$ Anomaly for Top:
$tZ'$ Production at LHC\footnote{
Talk presented at the APS Division of Particles and Fields Meeting
(DPF 2017), July 31-August 4, 2017, Fermilab. C170731}
}

\bigskip\bigskip

%+\addtocontents{toc}{{\it D. Reggiano}}
%+\label{ReggianoStart}

\begin{raggedright}

{\it George W.-S. Hou\footnote{
Home institute: Department of Physics, National Taiwan University,
Taipei 10617, Taiwan.} \\
ARC CoEPP at the Terascale \\
School of Physics, University of Melbourne\\
Melbourne, Vic 3010, AUSTRALIA
}
\bigskip\bigskip
\end{raggedright}

\section{Introduction}

The ``$P_5'$ anomaly'', reported~\cite{Aaij:2013qta,Aaij:2015oid} by the LHCb experiment
in angular analysis of $B \to K^*\mu^+\mu^-$ decay, has caught a lot of attention lately.
One intriguing interpretation is that the $C_9$ Wilson coefficient of
the electroweak penguin gets shifted by one unit from Standard Model (SM) expectation,
plausibly due to a heavy $Z'$ boson.
However, it is of some concern~\cite{Chang:2017wpl} that,
increasing data from 1~fb$^{-1}$~\cite{Aaij:2013qta} to 3~fb$^{-1}$~\cite{Aaij:2015oid},
the experimental significance did not improve. Furthermore,
a recent result~\cite{CMS:2017ivg} by the CMS experiment
based on 8 TeV data, is quite consistent with SM.
Thus, the issue would need the unfolding of LHC Run 2 data,
as well as turn-on of Belle II, to clarify.

The $\Delta C_9 \sim -1$ shift has motivated an extension~\cite{Altmannshofer:2014cfa}
of the so-called gauged $L_\mu - L_\tau$ model (muon number minus tau number),
by adding vector-like $Q$, $D$ and $U$ quarks that carry the extra U(1)$'$ charge,
where $Q$ stands for a left- and right- weak doublet while
$D$, $U$ are singlets, with obvious notation.
By mixing with $b$ and $s$ quarks, the exotic $Q$ and $D$ quarks
could generate the desired $\Delta C_9$, implying possible $t \to cZ'$ decay.
But since the tree-level $P_5'$ effect corresponds to SM at loop level in strength,
$t \to cZ'$ decay is rather suppressed.
Interestingly, the mixing of the $U$ quark with right-handed (RH) $t$ and $c$
are not constrained by $B$ physics, but even for this case,
$t \to cZ'$ decay is not too promising~\cite{FHK}.

Here, we point out~\cite{Hou:2017ozb} an alternative: searching for $cg \to tZ'$
generated by RH $tcZ'$ coupling.
Whether $P_5'$ anomaly in $B \to K^*\mu^+\mu^-$ turns out genuine or not,\footnote{
For a U(1) symmetry, one could have just the $U$ quark,
with $Q$ and $D$ absent.}
this would be akin to a ``$P_5'$ probe'' of the top quark itself.
We find good discovery potential in the coming future,
while the LH $tZ'$ that is related to $P_5'$ would be too weak
to observe in direct search at LHC.

\section{\boldmath RH $tcZ'$ and $ccZ'$ Couplings}

The $t_R$, $c_R$ quarks mix with $U_L$ by new Yukawa couplings
$Y_{Ut}$, $Y_{Uc}$ to an SM singlet scalar field $\Phi$,
which breaks the U(1)$'$ symmetry by $\langle \Phi\rangle = v_\Phi \neq 0$.
With $Z'$ emitted by the $U$ quark, one could consider 
an effective and therefore \emph{generic} coupling
\begin{equation}
 -\left( g^R_{ct}\, \bar{c}_{R}\gamma^\alpha t_{R} Z'_\alpha +{\rm h.c.}\right),
\end{equation}
where, specific to the extended $L_\mu - L_\tau$ model, one has
\begin{equation}
g^R_{ct}=\left( g^R_{tc}\right)^* =-g'\frac{Y^*_{Uc}Y_{Ut}v_{\Phi}^2}{2m_U^2},
\label{eq:gctR}
\end{equation}
as interpretation. Note that Eq.~(2) is nonzero only if $Y_{Uc}\neq 0$.
Thus, the model implies a RH $ccZ'$ coupling,
\begin{equation}
g^R_{cc}=-g'\frac{|Y_{Uc}|^2v_{\Phi}^2}{2m_U^2}, \label{eq:gccR}
\end{equation}
hence one could also contemplate a \emph{generic} coupling
\begin{equation}
 -g^R_{cc}\, \bar{c}_{R}\gamma^\alpha c_{R} Z'_\alpha.
\end{equation}

Eq.~(1) generates the $cg \to tZ'$ process,
where $Z' \to \mu^+\mu^-$ with 1/3 branching fraction.
We note that
\begin{equation}
\Gamma_{Z'} \simeq \frac{m_{Z'}^3}{4\pi v_\Phi^2}
  \simeq 0.75~{\rm GeV} \left( \frac{m_{Z'}}{150~{\rm GeV}} \right)^3
 \left( \frac{600~{\rm GeV}}{v_\Phi}\right)^2,
\end{equation}
is in general not long-lived, but also not too broad.

We follow closely a similar study by CMS using 8 TeV data
for $tZ$ final state~\cite{Sirunyan:2017kkr},
where the signal is $\ell\mu^+\mu^- + b\; +$ missing-$E_T$,
with backgrounds from $tZj$, $ttZ$, $ttW$, $WZ + j$, etc.
We start by considering two benchmark $m_{Z'}$ values,
one below $m_t$ and one above.
Events are generated by MadGraph and interfaced with PYTHIA for showering,
where Delphes is used for detector simulation.
After preselection cuts at generator level on
lepton $p_T$, $\eta$, separation, and jet $p_T$,
further selection cuts are applied for refined definition of
$\ell\mu^+\mu^-$, $b$-jet, lepton $p_T$,
top-related ``softest lepton'' and $W$-mass etc.,
and we employ a second $p_T$ jet veto to reduce $ttZ$ and $ttW$ background.
Effects of these cuts, such as plots and tables,
 can be found in Ref.~\cite{Hou:2017ozb}.
For signal significance, we use
$\mathcal{Z} = \sqrt{2 \left[ (S+B)\ln\left( 1+S/B \right)-S \right]}$.
Assuming $|g_{ct}^R| = 0.01$ and combining the $tZ'$ and $\bar tZ'$ processes,
we find 5$\sigma$ discovery potential for 180 (450) fb$^{-1}$ data at 14 TeV,
for $m_{Z'} = 150$ (200) GeV.
This ``discoverability'' motivates us to extend to full range of $m_{Z'}$
for the discovery reach in $|g_{ct}^R|$.

Eq.~(4) motivates ``milliweak'' Drell-Yan production of $Z'$
from $c\bar c$ parton annihilation, $c\bar c \to Z' \to \mu^+\mu^-$, which is interesting.
For our two benchmark points, we choose a slightly lower $|g_{cc}^R| = 0.005$ value.
%Since the width, Eq.~(5), is not too broad,
We find, through straightforward study and using
significance $\mathcal{Z} = S/\sqrt B$ as usual,
5$\sigma$ discovery potential for 110 (170) fb$^{-1}$ data at 14 TeV
for $m_{Z'} = 150$ (200) GeV, which seems promising.

%%%%%%%%%%%%%%%%%%%%%%%%%%%%%%%%%%%%%%%%%%%%%%%%%%%%%%%%%%%%%%%%%%%%%%%%%
%%
%%   use this format to include an .eps figure into your paper
%%
\begin{figure}[thb]
\begin{center}
\epsfig{file=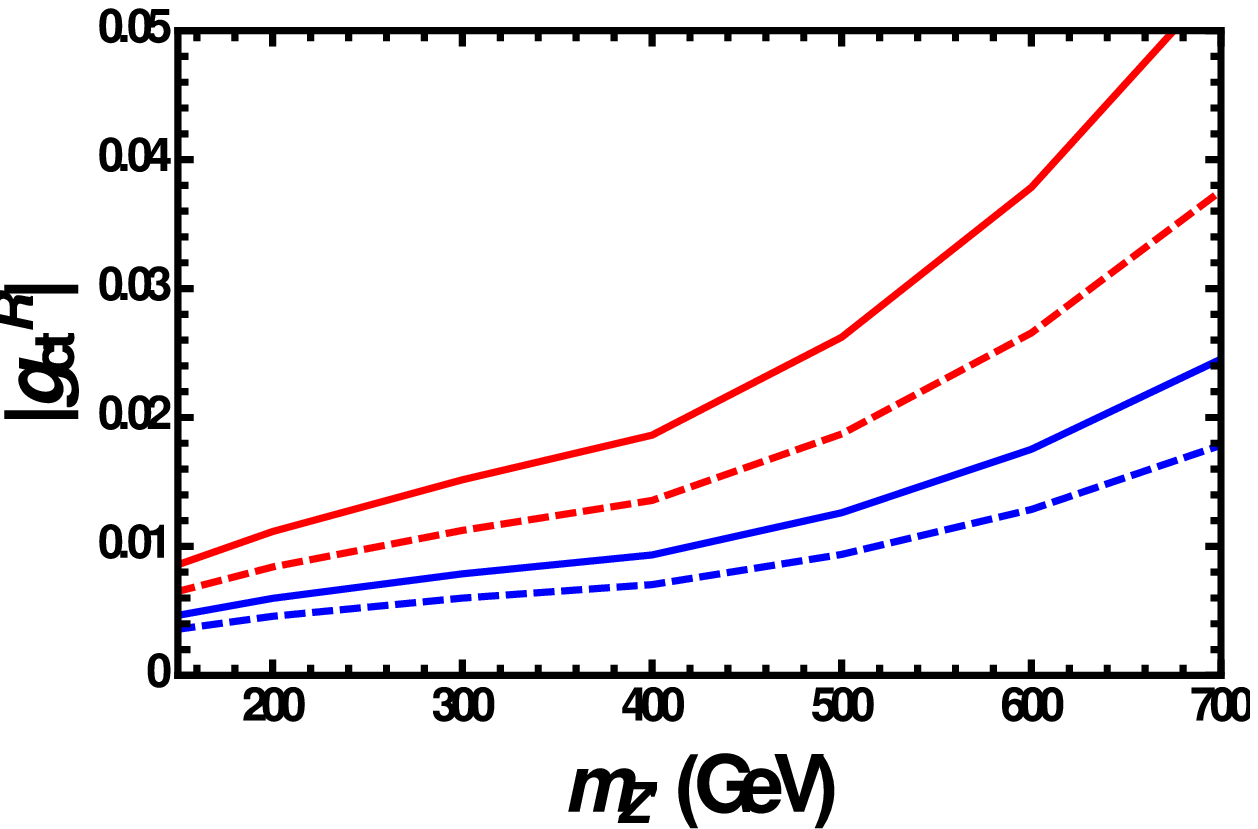,height=1.6in}
\epsfig{file=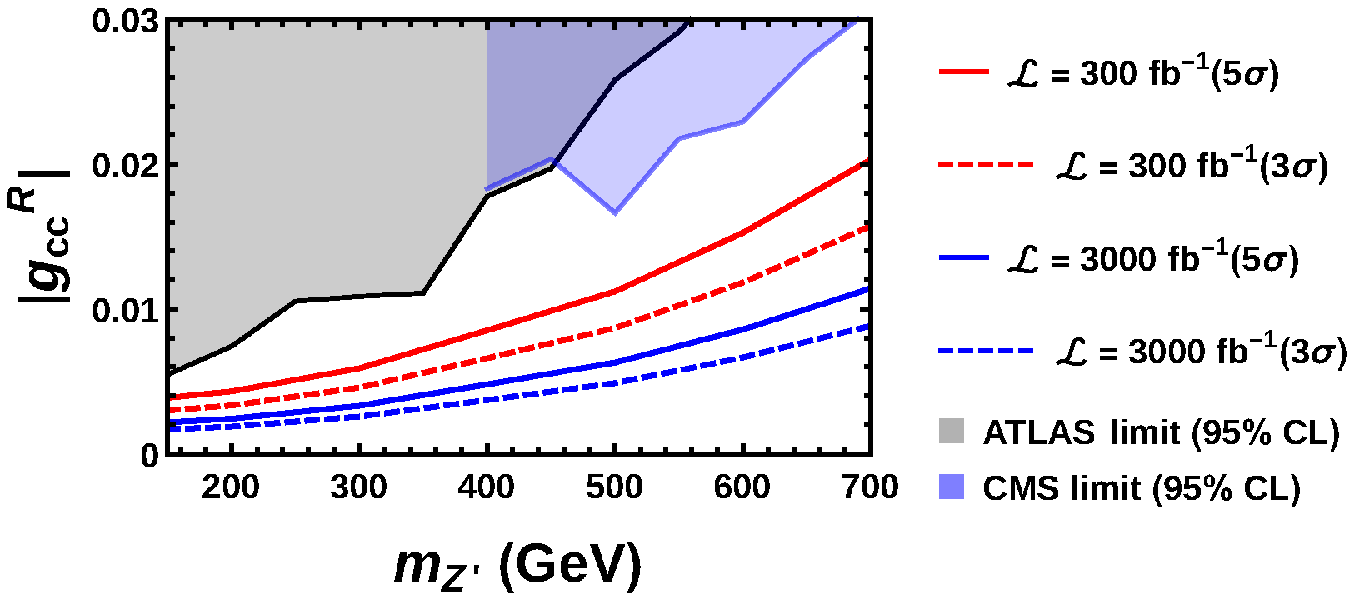,height=1.6in}
\caption{Discovery reach for [left] combined $pp \to tZ'$ and $\bar tZ'$,
   and [right] $pp \to Z' + X \to \mu^+\mu^- + X$ at 14 TeV LHC.}
%\label{fig:magnet}
\end{center}
\end{figure}
%%%%%%%%%%%%%%%%%%%%%%%%%%%%%%%%%%%%%%%%%%%%%%%%%%%%%%%%%%%%%%%%%%%%%%%%%%%

Extending the study for discovery reach, we plot
$|g_{ct}^R|$ vs $m_{Z'} \in (150,\, 700)$ GeV
in Fig.~1[left] for $tZ'$,
and $|g_{cc}^R|$ vs $m_{Z'}$ in Fig.~1[right] for dimuon.
For both figures, the upper (lower) solid curves are 5$\sigma$ discovery
for 300 (3000) fb$^{-1}$, while dashed curves stand for 3$\sigma$ evidence.
For Fig.~1[right], the preliminary dimuon search
limits of ATLAS~\cite{ATLAS:2016cyf} and CMS~\cite{CMS:2016abv},
each with $\sim$ 13 fb$^{-1}$ data at 13 TeV, are shown in the figure.
Not shown is the improved ATLAS 13 TeV result~\cite{Aaboud:2017buh}
with 36.1 fb$^{-1}$ data.

From the figures, we see that there is much parameter space
to be explored, hence discovery potential, for both $tZ'$ and dimuon processes.
For example, if we set the goal of coupling strengths of
$|g_{ct}^R|$ down to 0.025 or $|g_{cc}^R|$ down to 0.01,
then 300 (3000) fb$^{-1}$ data allows discovery up to 440 (640) GeV.
The lesson learned is that HL-LHC might still offer discoveries,
that of weaker-coupled (narrow) resonances.

\section{\boldmath Probing Gauged $L_\mu - L_\tau$}

Although the couplings $g_{ct}^R$ and $g_{cc}^R$ of Eqs. (1) and (4)
can be viewed as generic and independent,
let us give a possible physics interpretation of a potential,
future discovery at the LHC in the $L_\mu - L_\tau$ model.
We introduce the dimensionless ratio
\begin{equation}
 \delta_{Uq} \equiv {Y_{Uq} v_\Phi}/{\sqrt{2} m_U}, \quad\quad (q=t,\, c)
\end{equation}
which is the mixing angle between $U$ and $q = t,\; c$.
The yardstick then is the Cabibbo angle $\lambda \equiv \cos\theta_C \simeq 0.22$.
Given that the $U$ quark is quite exotic and heavy,
$\delta_{Ut}$ and $\delta_{Uc}$ might not be much larger than $\lambda$,
especially for $\delta_{Uc}$.

%%%%%%%%%%%%%%%%%%%%%%%%%%%%%%%%%%%%%%%%%%%%%%%%%%%%%%%%%%%%%%%%%%%%%%%%%
%%
%%   use this format to include an .eps figure into your paper
%%
\begin{figure}[thb]
\begin{center}
\epsfig{file=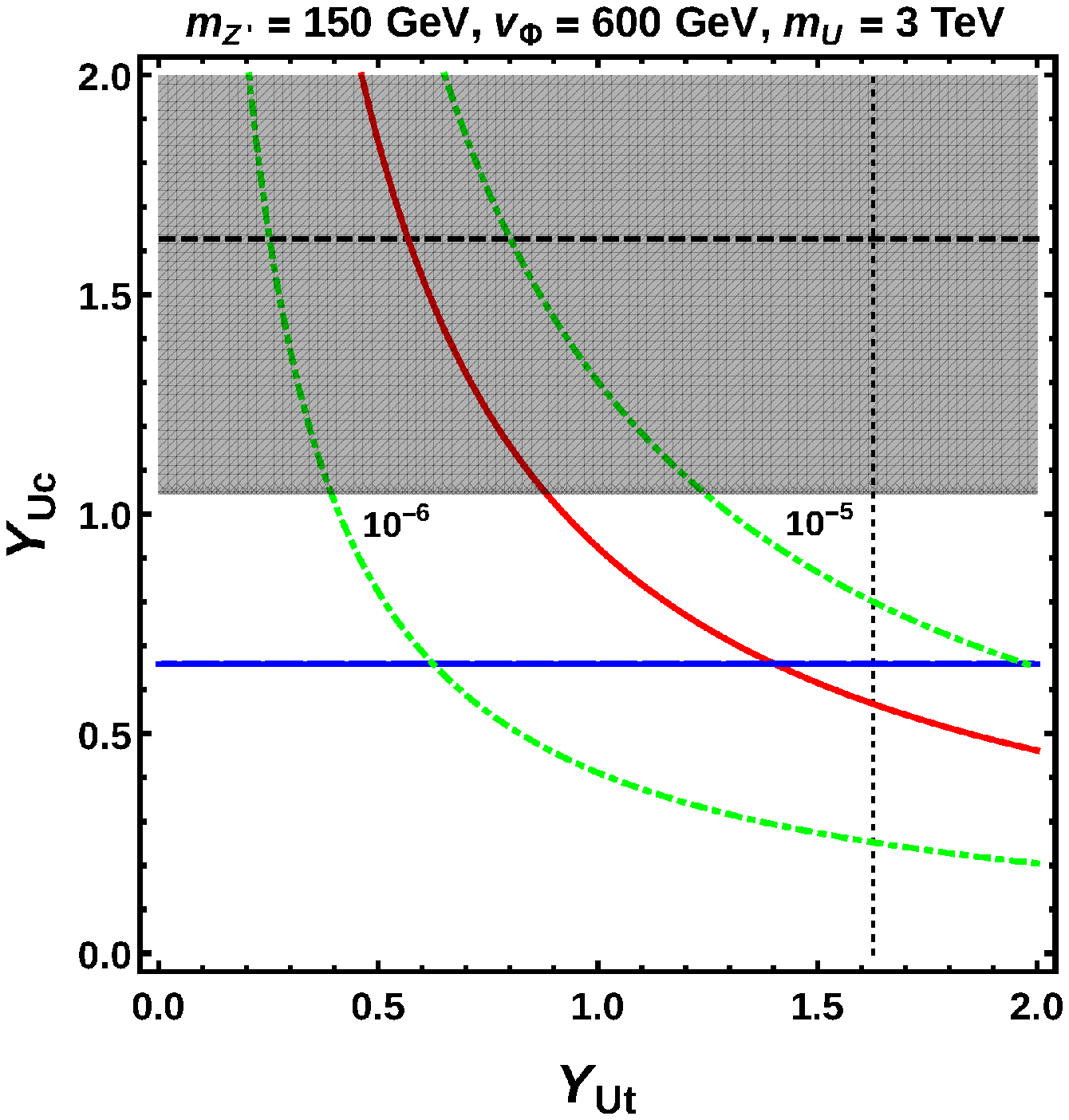,height=2.35in}
\epsfig{file=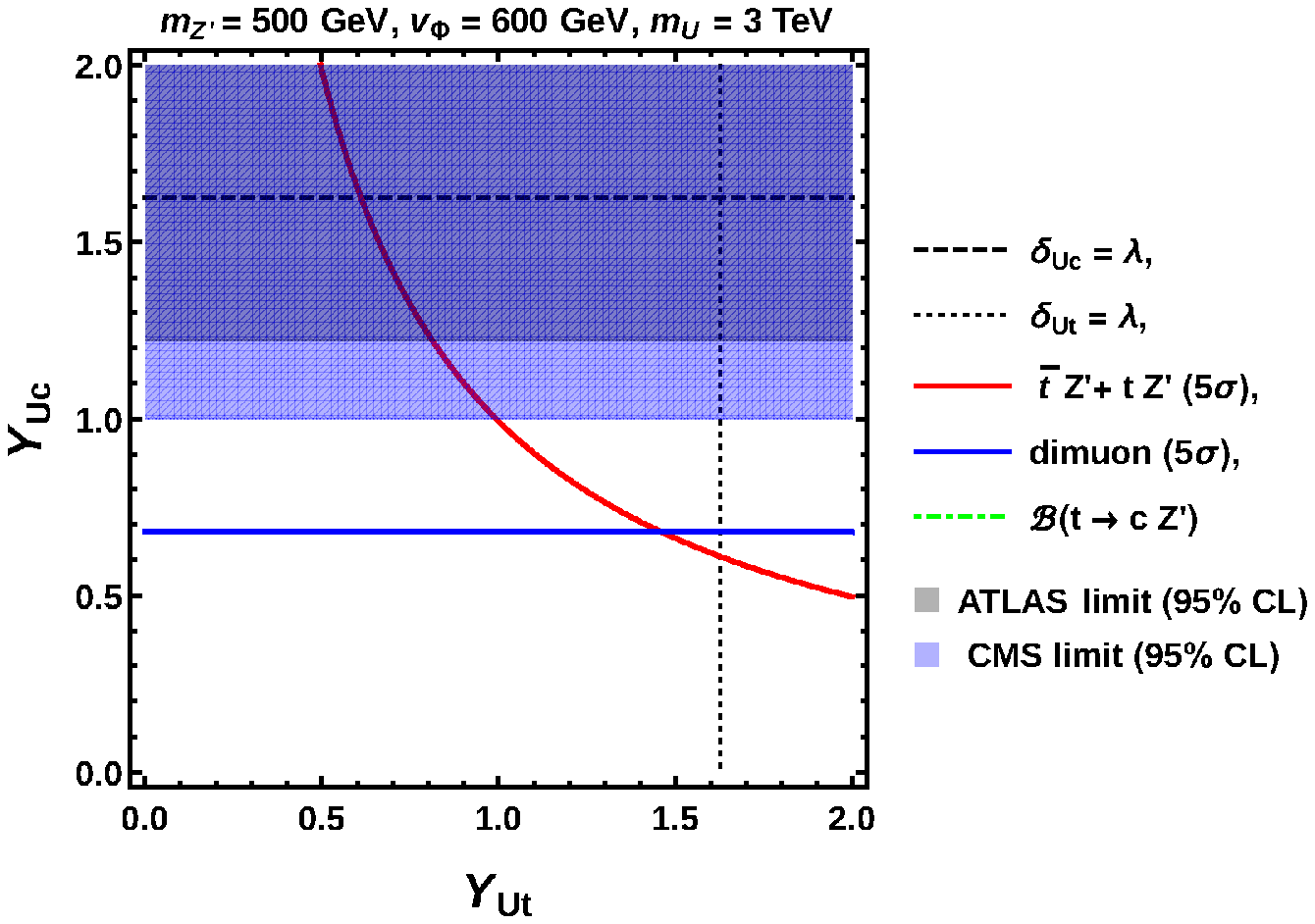,height=2.35in}
\caption{5$\sigma$ discovery reach in $Y_{Ut}$--$Y_{Uc}$ plane at 14 TeV LHC
   for [left] $m_{Z'} = 150$ GeV and [right] $m_{Z'} = 500$ GeV.
   Further explanations in text.}
%\label{fig:magnet}
\end{center}
\end{figure}
%%%%%%%%%%%%%%%%%%%%%%%%%%%%%%%%%%%%%%%%%%%%%%%%%%%%%%%%%%%%%%%%%%%%%%%%%%%

Motivated by low energy constraintes~\cite{FHK},
we set $v_\Phi = 600$ GeV, $m_U = 3$ TeV, and plot in Fig.~2
the 5$\sigma$ discovery reach in $Y_{Ut}$--$Y_{Uc}$ plane,
for $m_{Z'} =$ 150 GeV [left], 500 GeV [right],
where solid curve is for combined $tZ'$ and $\bar tZ'$,
and horizontal solid line is for dimuon.
For both figures, the horizontal dashed line marks $\delta_{Uc} = \lambda$
and vertical dotted line marks $\delta_{Ut} = \lambda$.
For Fig.~2[left], the case of $m_{Z'} = 150$ GeV which is below top threshold,
the two dash-dot lines are marked by
${\cal B}(t \to cZ') = 10^{-6},\; 10^{-5}$~\cite{FHK},
while the gray shaded region is excluded by ATLAS
preliminary dimuon search at 13 TeV~\cite{ATLAS:2016cyf} with $\sim 13$ fb$^{-1}$ data;
the 36 fb$^{-1}$ result~\cite{Aaboud:2017buh} by ATLAS at 13 TeV,
not shown, improves limit on $Y_{Uc}$ by $\sim 20\%$.
The CMS preliminary 13 TeV result~\cite{CMS:2016abv} does not probe this low $m_{Z'}$ value.
For Fig.~2[right], the case of $m_{Z'} = 500$ GeV which is considerably above top threshold,
we note from Eq.~(5) that $\Gamma_{Z'} \sim 27$ GeV,
and we have adjusted the $m_{\mu\mu}$ mass window to accommodate.
The region excluded by CMS preliminary dimuon search at 13 TeV with $\sim 13$ fb$^{-1}$ data
is somewhat better than ATLAS. However,
the 36 fb$^{-1}$ result~\cite{Aaboud:2017buh} by ATLAS at 13 TeV,
not shown, is comparable to the CMS preliminary result with smaller dataset.
It would therefore be interesting to see the CMS update with full 2015 and 2016 dataset.

%%%%%%%%%%%%%%%%%%%%%%%%%%%%%%%%%%%%%%%%%%%%%%%%%%%%%%%%%%%%%%%%%%%%%%%%%
%%
%%   use this format to include an .eps figure into your paper
%%
\begin{figure}[thb]
\begin{center}
\epsfig{file=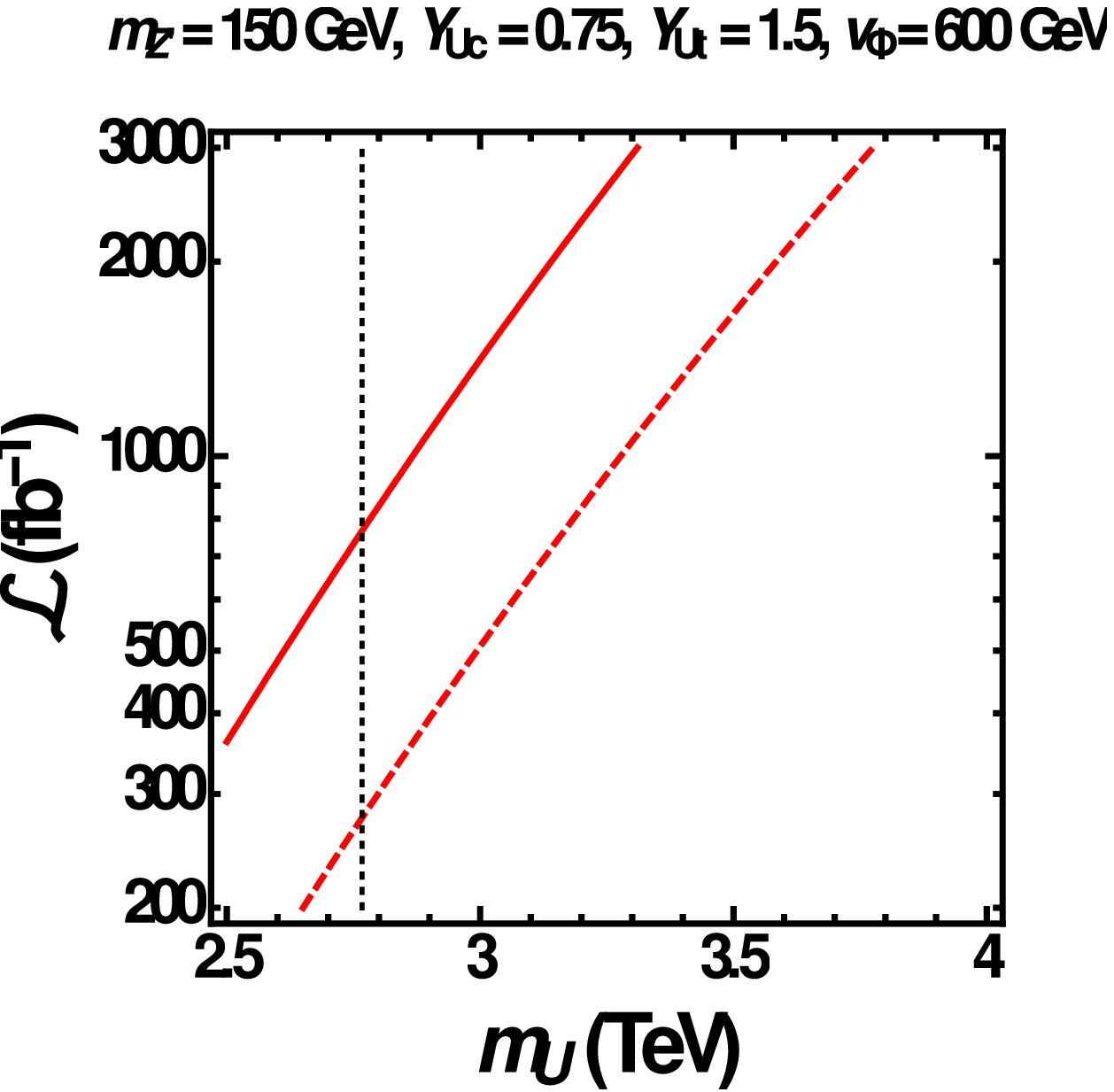,height=2.1in} \hskip0.5cm
\epsfig{file=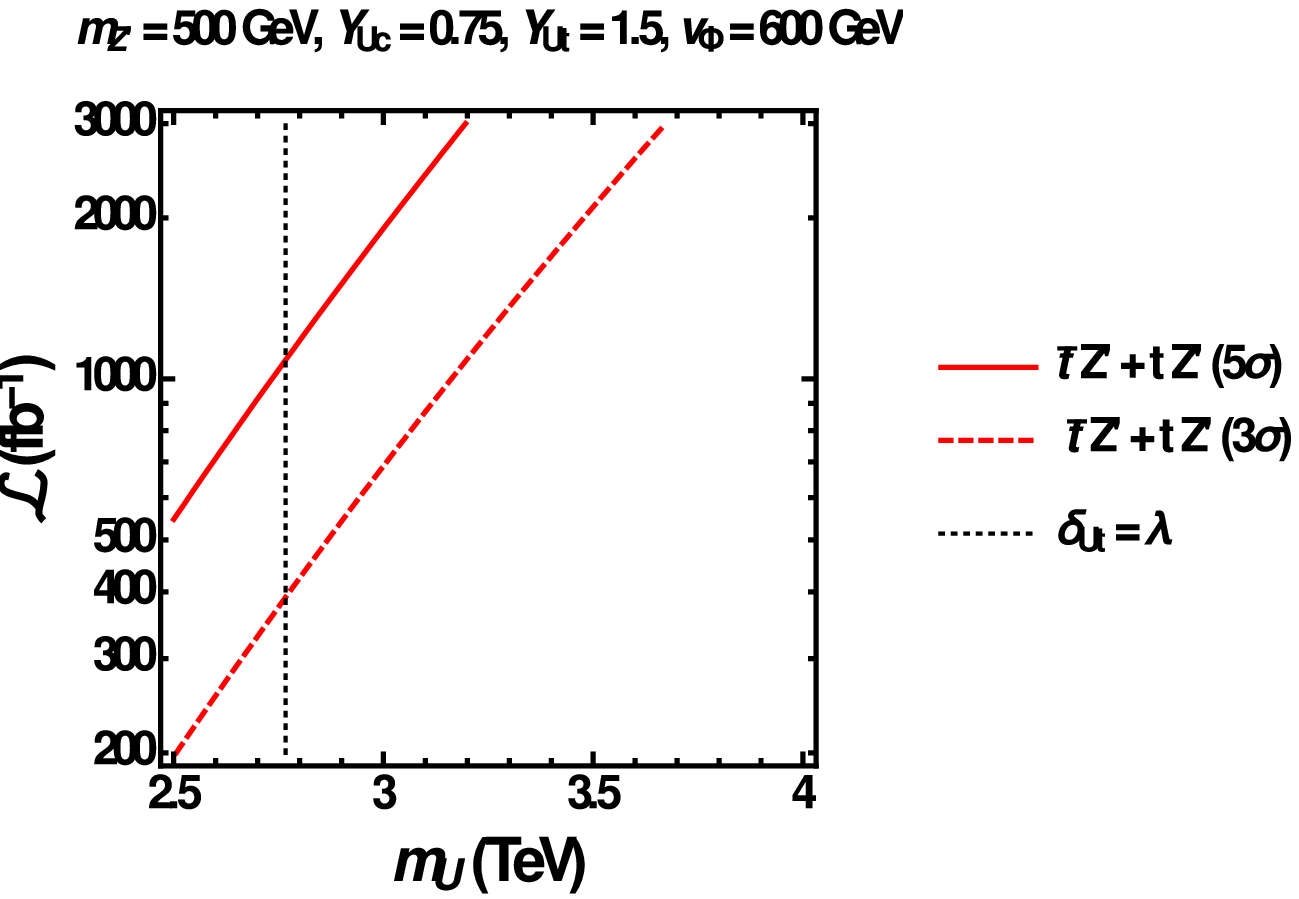,height=2.1in}
\caption{Discovery reach for combined $tZ'$ and $\bar tZ'$ vs $m_U$ at 14 TeV LHC
   for $m_{Z'} =$ 150 GeV [left], 500 GeV [right],
   for illustrative $Y_{Uc}$, $Y_{Ut}$ and $v_\Phi$ values.}
%\label{fig:magnet}
\end{center}
\end{figure}
%%%%%%%%%%%%%%%%%%%%%%%%%%%%%%%%%%%%%%%%%%%%%%%%%%%%%%%%%%%%%%%%%%%%%%%%%%%

A different perspective is to interpret the search as indirect probe of vector-like quark $U$.
To illustrate, we take $v_\Phi = 600$ GeV with a mild hierarchical pattern
for Yukawa couplings, $Y_{Ut} = 1.5$ and $Y_{Uc} = 0.75$.
Note the difference in parameter space from Fig.~2, as we allow $m_U$ to vary.
We plot integrated luminosities needed as a function of $m_U$
for 5$\sigma$ discovery (3$\sigma$ evidence) of the combined $tZ'$ and $\bar tZ'$ process
as solid (dashed) lines in Fig.~3, for $m_{Z'} =$ 150 GeV [left], 500 GeV [right].
Again, the vertical dotted line marks $\delta_{Ut} = \lambda \simeq 0.22$,
decreasing to the right as $m_U$ increases (see Eq.~(6)).
We see that a discovery, which could happen with $\sim$~300 fb$^{-1}$ or more data
(and indication for $m_{Z'} = 150$ GeV case could appear within Run 2),
one would be probing $m_U$ above 2.5 TeV.
As higher $m_U$ is probed, one probes smaller $\delta_{Ut}$ mixing
as allowed by the higher integrated luminosity,
which is indeed analogous with $P_5'$-like sensitivities with number of $B$ mesons.

%%%%%%%%%%%%%%%%%%%%%%%%%%%%%%%%%%%%%%%%%%%%%%%%%%%%%%%%%%%%%%%%%%%%%%%%%
%%
%%   use this format to include an .eps figure into your paper
%%
\begin{figure}[t]
\begin{center}
\epsfig{file=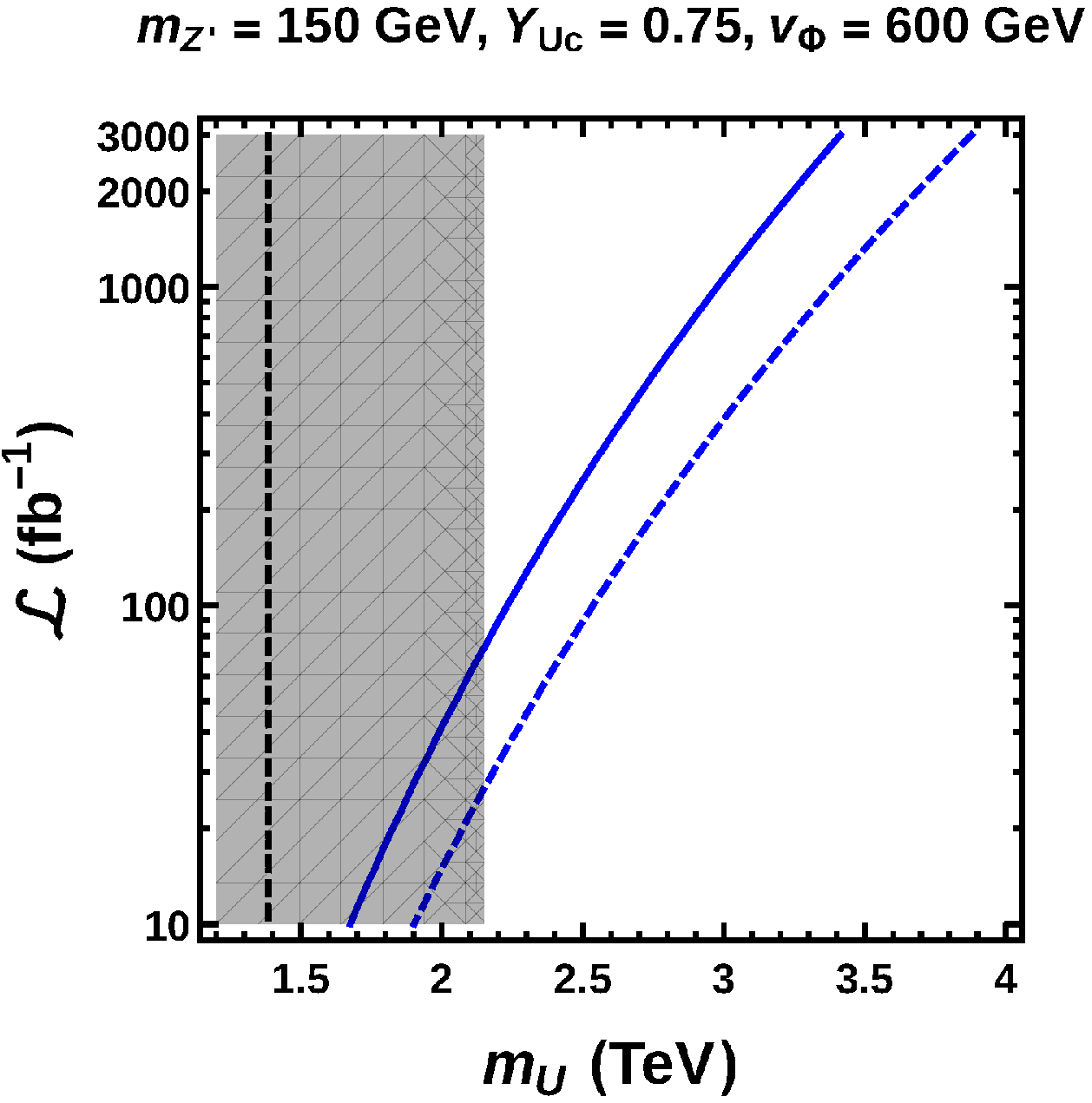,height=2.15in} \hskip0.2cm
\epsfig{file=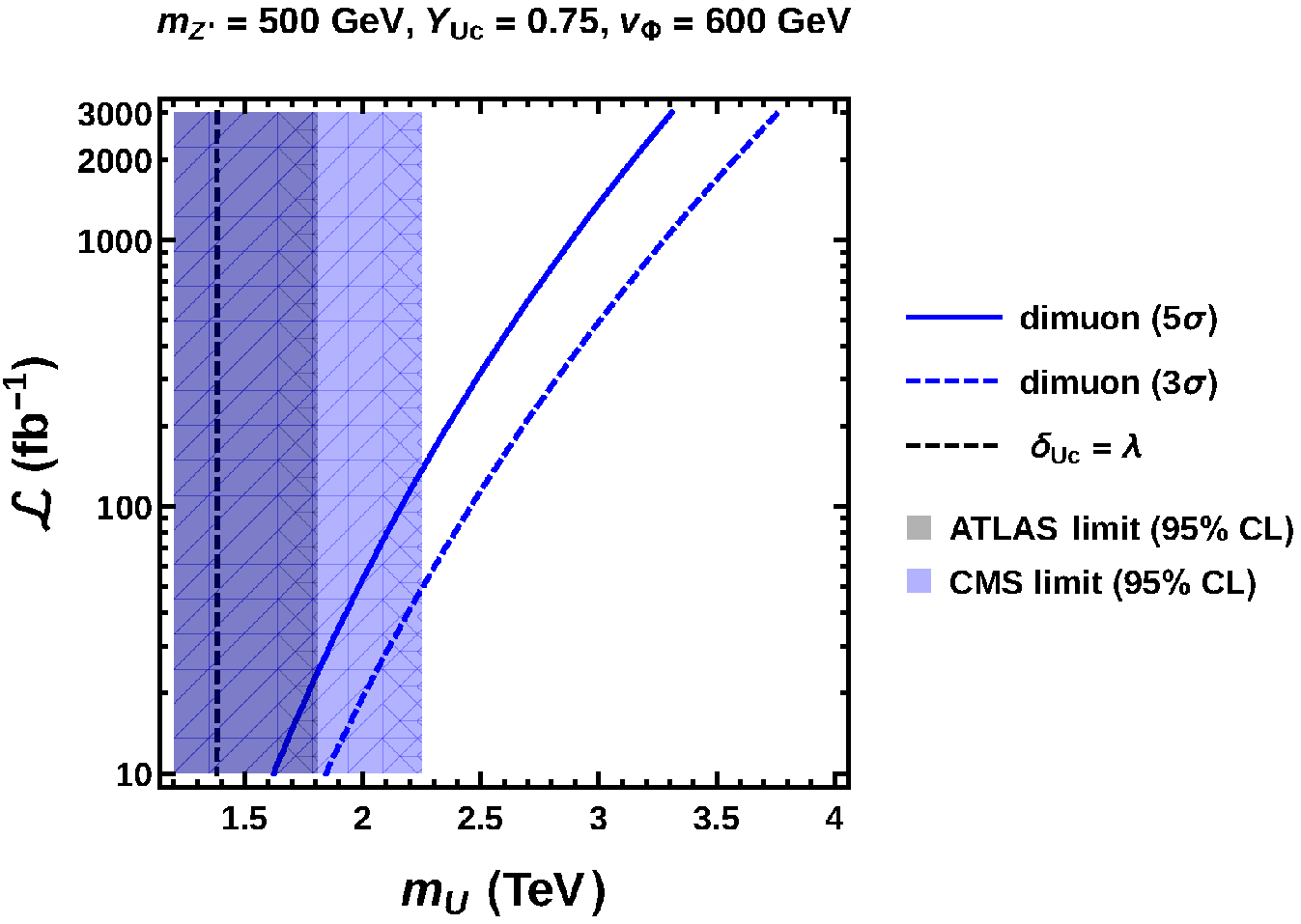,height=2.15in}
\caption{Same as Fig.~3 but for dimuon search.}
%\label{fig:magnet}
\end{center}
\end{figure}
%%%%%%%%%%%%%%%%%%%%%%%%%%%%%%%%%%%%%%%%%%%%%%%%%%%%%%%%%%%%%%%%%%%%%%%%%%%

We give the analogous plots, with $Y_{Ut}$ now unspecified,
for dimuon process in Fig.~4, where now the vertical dashed line
marks $\delta_{Uc} = \lambda \simeq 0.22$, decreasing to the right as $m_U$ increases.
We see that the dimuon search at 13 TeV (shaded regions) by
ATLAS and CMS are already probing $m_U$ above 2 TeV,
and higher luminosities allow one to probe smaller $\delta_{Uc}$ mixings
for heavier $m_U$.

\section{Discussion and Conclusion}

If mixing angles $\delta_{Uc} \sim \delta_{Ut}$, then 
dimuons would be more promising.
If discovered, one should study $cg \to cZ'$ with
the aid of $c$-tagging, if feasible.
For the hierarchical pattern and $\delta_{Uc}/\delta_{Ut} < 0.4$,
then $tZ' + \bar tZ'$ becomes more promising.
If discovered, then the handedness can be studied via angular analysis.
Both topics are under investigation.
If both $tZ'$ and dimuons are discovered,
we could learn the detailed flavor structure.
However, if one takes the prejudice that
$\delta_{Ut} < \lambda \sim 0.22$ and $\delta_{Uc}/\delta_{Ut} < \lambda$,
then the scenario seems out of reach at the LHC.
But experimental search should certainly be pursued,
as one can consider these as generic processes.

We note that, for LH $g_{ct}^L$ coupling that is directly linked to
$B \to K^{(*)}$ anomalies (not just $P_5'$),
$tZ'$ would be out of reach even for the HL-LHC, because
the effective coupling at the SM loop level makes it too weak to be produced.
But maybe one also should contemplate generic $b\bar b \to Z'$?
This would enter the $cg \to cZ'$ program.

In conclusion,
we have illustrated the possible emergence of ``Top Flavor Physics''
with $tZ'$ and $Z' \to \mu^+\mu^-$ dimuon search at the LHC.
The discoverability of this ``$P_5'$ of top''
could start with Run 2, all the way to the end of HL-LHC.
A moral can be drawn from this:
weaker-coupled resonances are probed at Run 2, and by HL-LHC.
We should not give up hope for discovery.

\bigskip
We thank Masaya Kohda and Tanmoy Modak for pleasant collaboration.

\end{document}